\documentclass[conference]{IEEEtran}

\usepackage{cite}
\usepackage{amsmath,amssymb,amsfonts}
\usepackage{graphicx}
\usepackage{textcomp}
\usepackage{xcolor}
\usepackage[noend]{algpseudocode}
\usepackage{algorithmicx,algorithm}
\newcommand{\eg}{\textit{e}.\textit{g}.}

\newcommand{\ie}{\textit{i}.\textit{e}.}
\newcommand{\etc}{\textit{etc}}

\usepackage{caption}
\usepackage{booktabs}
\begin{document}
\title{MYCloth: Towards Intelligent and Interactive Online T-Shirt Customization based on User's Preference}
\author{\IEEEauthorblockN{Yexin Liu}
\IEEEauthorblockA{AI Thrust, HKUST(GZ)\\
Email: yliu292@connect.hkust-gz.edu.cn}
\and
\IEEEauthorblockN{Lin Wang$^\dagger$}
\IEEEauthorblockA{AI/CMA Thrust, HKUST(GZ) and\\Dept. of CSE, HKUST\\
Email: linwang@ust.hk}}

\maketitle

\begin{abstract}
In conventional online T-shirt customization, consumers, \ie, users, can achieve the intended design only after repeated adjustments of the design prototypes presented by sellers in online dialogues. However, this process is prone to limited visual feedback and cumbersome communication, 
thus detracting from users' customization experience and time.
This paper presents an intelligent and interactive online customization system, named \textbf{MYCloth}, aiming to enhance the T-shirt customization experience. Given the user's text input, our MYCloth employs ChatGPT to refine the text prompt and generate the intended paint of the cloth via the Stable Diffusion model. Our MYCloth also enables the user to preview the final outcome via a novel learning-based virtual try-on model. The whole system allows to iteratively adjust the cloth till optimal design is achieved.
We verify the system's efficacy through a series of performance evaluations and user studies, highlighting its ability to streamline the online customization process and improve overall satisfaction.
\end{abstract}

\IEEEpeerreviewmaketitle

\section{Introduction}
Traditionally, obtaining T-shirts involves physical store visits or more recently, online shopping. Consumers, \ie, users, seek out pre-made clothes that match their personal style. The clothing industry responds by providing a wide range of fashion items for various tastes and occasions~\cite{zhao2018manufacturing,matthews2019impact,de2019collaborative}. However, these options are limited by mass production and standard designs. A growing trend focuses on personalized T-shirts, where individuals can express their unique style, turning their attire into a form of self-expression.
The essence of personalized T-shirts lies in their capacity to offer consumers an active role in the design and conceptualization of their clothing. It is a journey where consumers transit from passive wearers to active participants in its creation. Through a blend of innovative technology and collaborative design processes, individuals are empowered to make choices concerning fabric, color, patterns, and even the minutiae of stitching details.

Recently, online customization has gained great attention as it allows customers to design and order clothes from the comfort of their homes or anywhere with internet access~\cite{manko2023evolution,nobile2021review,huang2021now,tudjarov2008web}. 
\begin{figure}[t!]
    \centering
    \includegraphics[width=.5\textwidth]{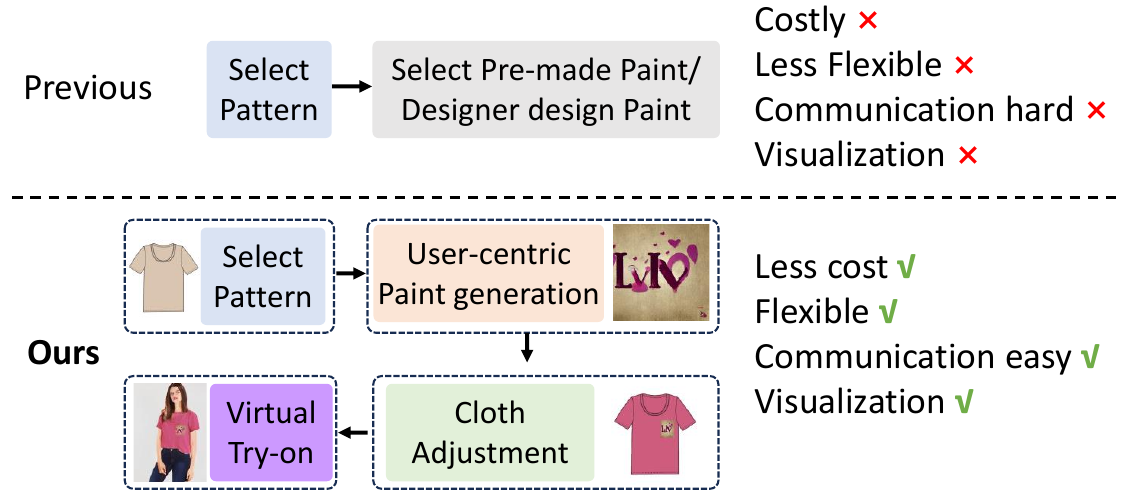}
    \caption{Comparison between previous T-shirt customization and our proposed AI-assisted personalized customization.}
    \label{paint_generation}
    \vspace{-15pt}
\end{figure}
 However, it also comes with a set of challenges:
\textbf{1) Communication Complexity}. Traditional systems for customized clothing often follow a linear path, with users initially conveying their specific design requirements to sellers. The burden of interpreting these requirements and crafting design prototypes rests primarily with the sellers. This approach, while common, can lead to communication complexities, misunderstandings, and protracted interactions.
\textbf{2) Limited Visual Feedback}. In many personalized clothing processes, there's a lack of effective tools for users to gain an immediate and intuitive visual understanding of how their chosen designs look on them. This absence of direct visual feedback can lead to uncertainty and dissatisfaction among users.

To address these challenges, some systems have been developed for interactive T-shirt design~\cite{teespring,zazzle}, enabling users to rearrange the clothing pattern elements, choose the available colors, and fabric. 
However, these systems 1) only allow users to select existing elements in the system during design, such as clothing texture, layout, \etc., and cannot generate some elements according to the user's own preferences, such as printing; 2) only allow users to view the garment itself without previewing how it looks when worn by a person. 

To this end, this paper proposes an intelligent and interactive
 online customization system, named \textbf{MYCloth}, aiming to enhance the T-shirt customization experience. Through MYCloth, it is possible to empower users to seamlessly select T-shirt styles, customize colors, and explore diverse design aesthetics, while offering interactive tools for comprehensive T-shirt customization. Technically, given the user's text input about the design theme,
 our MYCloth system first refines the text prompt by employing large language models (LLMs), such as ChatGPT~\cite{chatgpt}, to re-prompt the text to make it more consistent with the design theme. We then leverage the artificial intelligence (AI)-driven text-to-image generation model, \eg, Stable Diffusion model~\cite{rombach2022high}, to generate paint images that can be used in T-shirt design. Moreover, we propose a novel AI-based virtual try-on model that allows users to preview their customized T-shirts comprehensively. This enables individuals to gain valuable insights into the final appearance of their T-shirt and make necessary adjustments. In summary, the major contributions and novel aspects of our work are as follows:

\begin{itemize}
  \item \textbf{Novel System.} We introduce the MYCloth system, an intelligent and interactive platform for online T-shirt customization. MYCloth empowers users to seamlessly design and personalize their T-shirts by offering intuitive tools for style selection, color choices, and design customization. 
  \item \textbf{Novel Algorithm.} 1) \textbf{A LLM-assisted Paint Generation Model}. Our algorithm presents a novel approach by integrating ChatGPT to refine and specify user-provided textual descriptions. This enhancement results in more detailed and accurate inputs for our text-to-image generation model, ultimately leading to higher-quality, personalized clothing designs. 2) \textbf{A Novel Learning-based Virtual Try-On Model}. We propose an advanced virtual try-on model that leverages state-of-the-art computer vision and deep learning techniques. This model enables users to visualize and interact with their customized T-shirts on virtual avatars, providing a realistic and immersive online try-on experience.
  \item \textbf{Comprehensive Evaluation.} 1) Numerical evaluation. 2) Visual comparison; 3) User study. Through quantitative metrics and qualitative feedback, we demonstrate the effectiveness and efficiency of the MYCloth system.
\end{itemize}

\begin{figure}[t!]
    \centering
    \includegraphics[width=.5\textwidth]{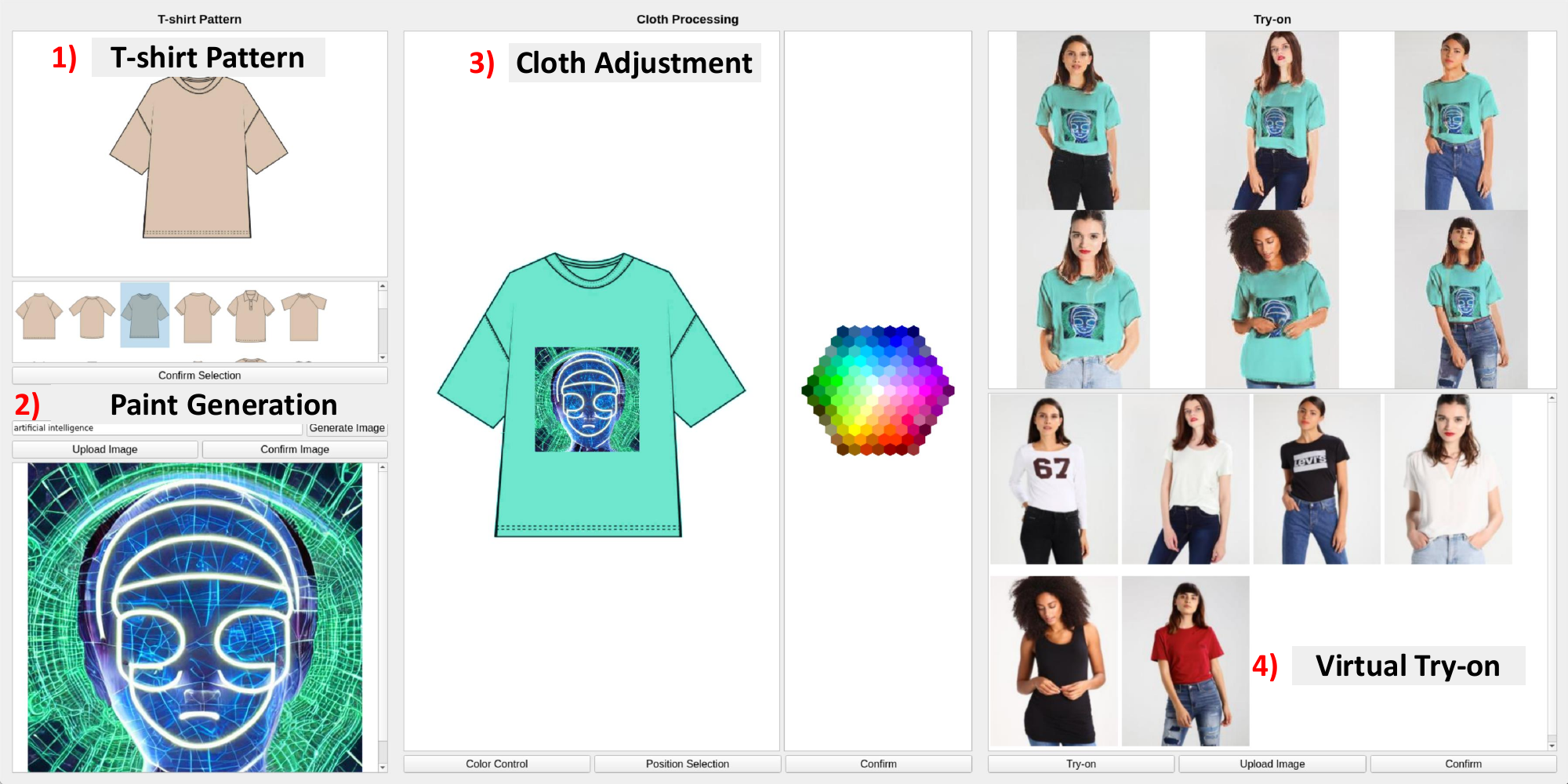}
    \caption{Overview of the proposed system: 1) \textbf{Pattern selection} allowing users to choose from a variety of T-shirt patterns, 2) \textbf{Paint Generation} intelligently generating design concepts aligned with user-defined themes, 3) \textbf{Cloth Adjustment} empowering users to meticulously refine design elements like color and the position of the paints, bridging the conceptualization and tangible design, and 4) \textbf{Virtual Try-On} enabling users to visualize their creations on 2D avatars for design validation.}
    \label{paint_generation}
    \vspace{-15pt}
\end{figure}
\section{Related Work}

\subsection{Computer-aided Clothing Design}
The intersection of artificial intelligence (AI) and fashion design, notable for its academic and industrial significance~\cite{zhu2022application,li2022application,sarmakari2022just,wu2023styleme}, has led to significant advancements in apparel Computer-Aided Design (CAD) systems, fabric and color selection, fit optimization, image synthesis, and the evaluation of virtual 3D designs. Systems like StyleMe~\cite{wu2023styleme} exemplify AI's role in automating design generation. Despite these developments, traditional design tools still heavily rely on the expertise of skilled designers, limiting inclusivity for non-experts. Interactive design systems~\cite{zhu2018interactive,zhu2020interactive,wang2022knowledge} allow for pattern and fabric customization, but their utility in broader aspects of fashion customization, such as T-shirt design where customers focus on colors and prints rather than patterns, remains limited.

\subsection{Text-to-image Generation}

Text-to-image generation models, like TediGAN~\cite{xia2021tedigan} and DALL·E~\cite{ramesh2021zero}, use textual descriptions to create realistic images, bridging linguistic creativity and visual design~\cite{esser2021taming,ding2021cogview,zhang2022armani}. These models, applicable in fashion design, fall into two categories: GAN-based methods (e.g., TediGAN using StyleGAN~\cite{karras2019style}) and two-stage methods. The latter involves transforming images into discrete tokens via variational autoencoders and then reconstructing them into images, as seen in DALL·E and Cogview~\cite{ramesh2021zero,ding2021cogview}. ARMANI enhances this process by integrating textual information early on for more expressive tokens~\cite{zhang2022armani}. While these models expand fashion visuals, they lack the ability for detailed garment customization, like specific print incorporation. This paper diverges from generating garments via text-to-image models, focusing instead on a re-prompting method based on large language models, such as ChatGPT, coupled with stable diffusion techniques for generating specific designs~\cite{chatgpt,rombach2022high}.

\subsection{AI-driven Virtual Try-on}
Virtual try-on technologies, crucial in digital garment visualization, utilize a two-stage process involving garment deformation and image synthesis~\cite{baldrati2023multimodal,xie2023gp,yan2023linking,zhu2023tryondiffusion,huang2022towards,yang2022full,dong2022dressing,ge2021parser,lee2022high}. Previous methods focused on neural network-based control points for deformation~\cite{lee2022high,issenhuth2020not} or appearance flow maps for non-rigid deformation~\cite{bai2022single,issenhuth2020not}. This paper proposes a novel one-stage framework for more efficient virtual try-on generation.

\section{System design and Implementation}
\begin{figure*}[t!]
    \centering
    \includegraphics[width=\textwidth]{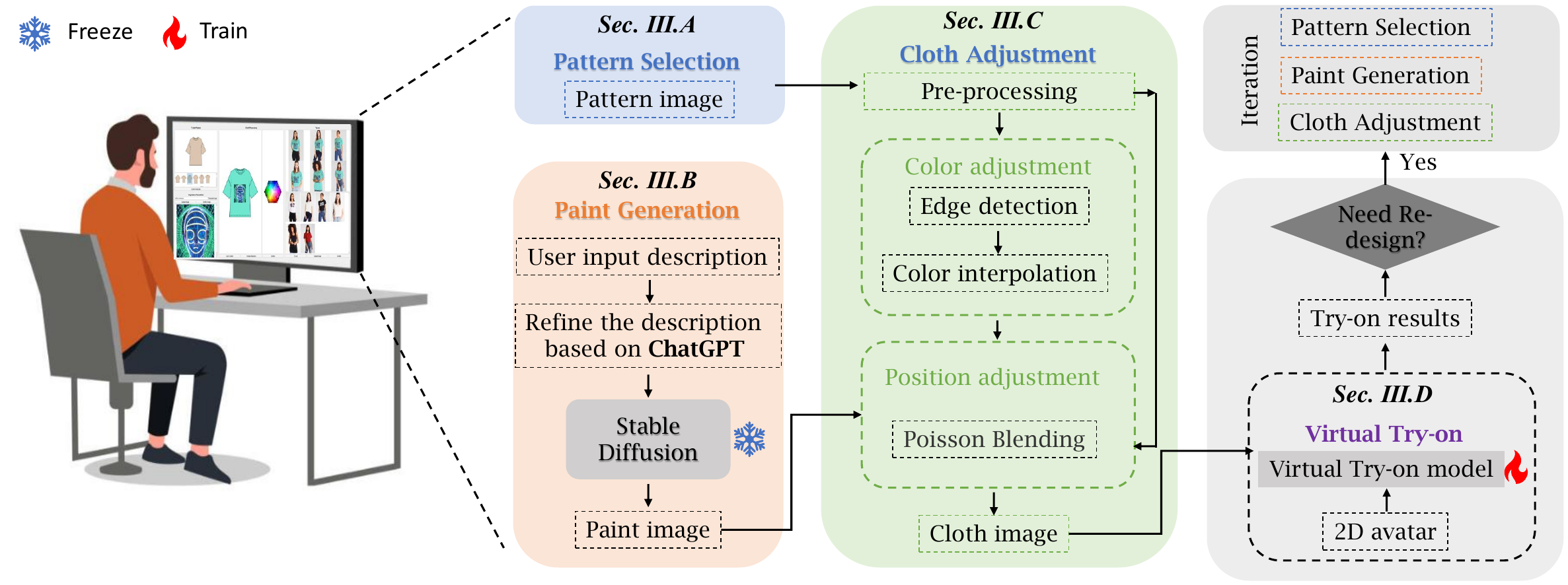}
    \caption{The overall structure of our system.}
    \label{framework_tech}
\end{figure*}

In this section, we introduce the proposed intelligent and interactive online T-shirt customization system, named MYCloth, tailored to provide users (\ie, consumers) with an engaging and efficient T-shirt customization experience. 
The overall system is shown in Fig.~\ref{framework_tech}.
The MYCloth system is composed of four key components: Pattern Selection, Paint Generation, Cloth Adjustment, and Virtual Try-On. Designed to simplify the creative customization process, users can seamlessly design T-shirts through intuitive interactions, such as clicking buttons and drag-and-drop actions. The sections are structured as follows: Section~\ref{Pattern Selection Implementation} illustrates the details of the pattern selection. Subsequently, Section~\ref{paint Generation Implementation} introduces a novel method for generating paints in accordance with user preferences. The subsequent Section~\ref{Cloth Adjustment} is dedicated to the image processing methods of cloth adjustment. Section~\ref{Virtual Try-On}, introduces a robust learning-based virtual try-on model that vividly visualizes the designs on virtual avatars.

\subsection{Pattern Selection}
\label{Pattern Selection Implementation}
The pattern selection phase serves as a fundamental starting point within the system. 
In this phase, customers are presented with a collection of T-shirt patterns. These patterns have been meticulously designed by experienced fashion designers. This collaboration ensures that those customers can access high-quality patterns that serve as a basis for their creative exploration. 
In essence, the pattern selection embodies MYCloth's commitment to inclusivity and collaboration. Through this approach, we aim to provide users with a solid foundation for their design journey, regardless of their experience level. 

\subsection{LLM-assisted Paint Generation Model}
\begin{figure}[t!]
    \centering
    \includegraphics[width=.5\textwidth]{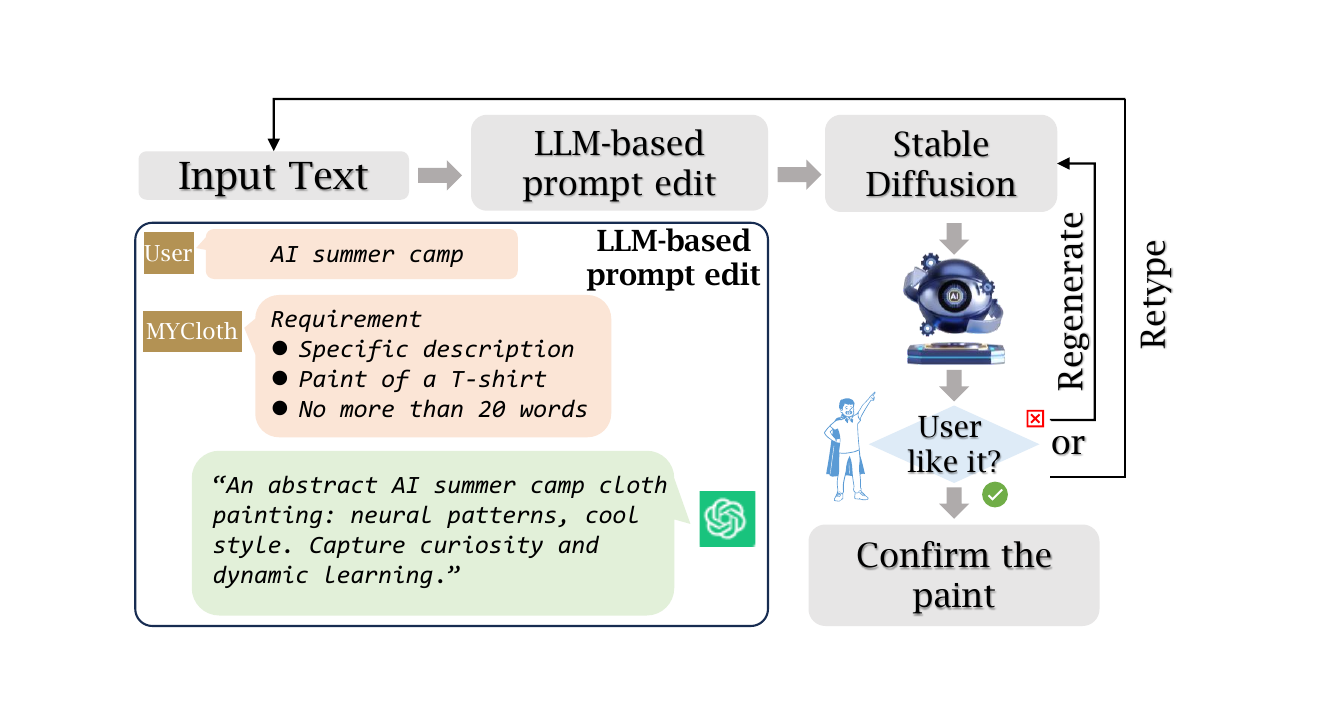}
    \caption{The overview of the paint generation. Users input text, which is optimized using ChatGPT, and the stable diffusion model is employed to generate prints.}
    \label{paint_generation}
    \vspace{-15pt}
\end{figure}

\label{paint Generation Implementation}

Customizing prints currently involves three methods: 1) Consumers design the print themselves, 2) they select from existing designs offered by sellers, or 3) they propose a theme and have professional illustrators design it, which incurs additional costs and time. This paper explores using a text-to-image generation model to create paint based on textual descriptions of design themes. However, slight variations in descriptions can lead to significant differences in outcomes. To address this, we propose a two-step solution: first, refining the textual description using ChatGPT, and then feeding these enhanced descriptions into a text-to-image model employing stable diffusion techniques for paint generation, as depicted in Fig~\ref{paint_generation}. This process starts with the user's theme description, which is refined through interaction with ChatGPT, and then used in the text-to-image process to create the paint. This method combines creative input with AI, producing paint that accurately reflects the intended theme.

\subsection{Cloth Adjustment}
\label{Cloth Adjustment}
The cloth adjustment component allows users to easily tailor and refine T-shirt designs to their liking. It offers an intuitive interface for adjusting elements like color, as well as the position and scale of the paint, thus enhancing the customization and personalization of T-shirts. This component comprises two main features: color adjustment and position adjustment.

\textbf{1) Color adjustment}. 
The cloth image processing involves several steps to maintain visual integrity and color consistency. Initially, pre-processing techniques denoise the image and analyze the pixel distribution. Edge detection algorithms are then applied to accurately identify and preserve the cloth's boundary colors and positions. Following this, a pixel interpolation method is used to ascertain the color information for each pixel. This pre-processing phase gathers the cloth's overall color distribution, recording the main image color as ($R_m, G_m, B_m$). For each pixel, the process determines whether it lies on the cloth's edge; if so, its color remains unchanged. Otherwise, the final pixel color is calculated through interpolation, considering its original color, the cloth's main color ($R_m, G_m, B_m$), and the target cloth color ($R_t, G_t, B_t$).

\textbf{2) Position adjustment}. The position adjustment function enables users to precisely control the placement of the paint image within the interface. By clicking and dragging within the boundaries of the paint, users can freely relocate it to any desired position and scale within the cloth. 

\subsection{Learning-based Virtual Try-On Model}
\label{Virtual Try-On}
\begin{figure}[t!]
    \centering
    \includegraphics[width=.5\textwidth]{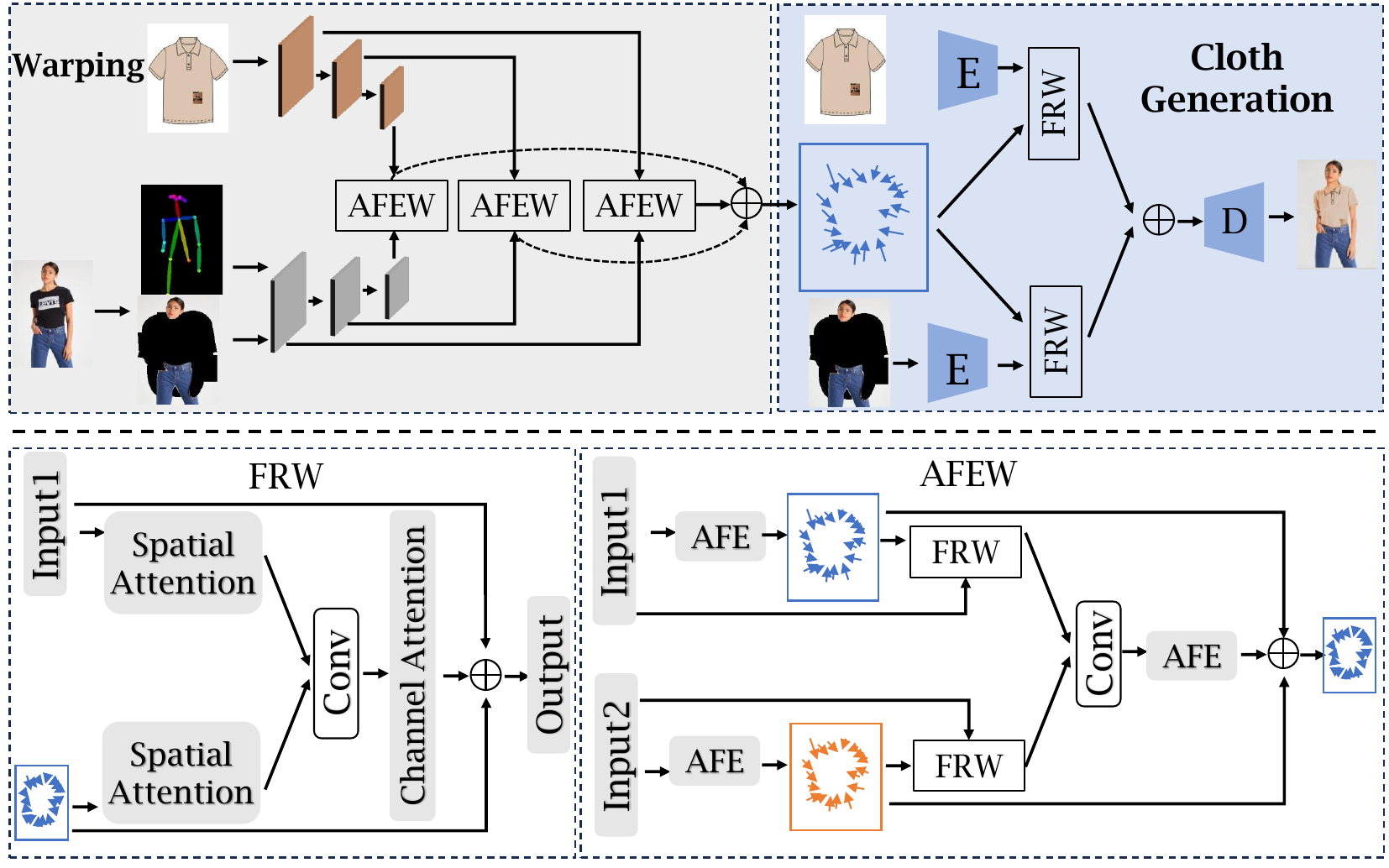}
    \caption{The virtual try-on framework. \textbf{AFEW}: attention-based flow estimation and warping module. \textbf{FRW}: flaw rectification and warping block. \textbf{AFE}: appearance flow estimator}
    \label{try_on}
    \vspace{-15pt}
\end{figure}

Our system integrates a virtual try-on, allowing users to project their T-shirt designs onto virtual avatars. This offers a seamless transition from digital design concepts to practical visuals, aiding users in making more informed design choices. As depicted in the figure, the system employs a single-stage framework for generating virtual try-on results. This framework comprises two key components: a warping module and a cloth generation module.

\subsubsection{Warping Module}

The inputs comprise the cloth image ($x_c$) and the 2D human image ($x_h$). The cloth image undergoes processing through the cloth branch, while the 2D human images are initially subjected to preprocessing to generate the human pose image and the human agnostic image. The latter masks the upper body of the depicted person. Subsequently, the human image branch receives the concatenated form of the human pose image and the human agnostic image. Both the cloth and the human branch have the same feature extraction structure, namely the Feature Pyramid Network (FPN)~\cite{lin2017feature}. Notably, the two branches do not share the parameters. For each scale, individual feature maps ($f_c^i$ and $f_h^i$) representing the cloth and human branches are obtained.

To facilitate the transformation of the cloth image to align with the contours of the human body, a methodology inspired by~\cite{ge2021parser,guler2018densepose,bai2022single} is adopted to entail the estimation of appearance flow across scales.  Specifically, predicated upon the extracted features $f_c^i$ and $f_h^i$ for each scale, an attention-based flow estimation and warping module (AFEW) is introduced, as depicted in Fig.\ref{try_on} (b). The appearance flow estimator (AFE)\cite{bai2022single} is introduced to deduce flow vectors $F_c^i$ and $F_h^i$. The formulation is as follows.

\begin{equation}
\begin{gathered}
F_c^i=AFE(f_c^i)\\
F_h^i=AFE(f_h^i)
\label{eq:TTA}
\end{gathered}
\end{equation}

To refine the acquired appearance flow, a flaw rectification and warping block (FRW) is proposed. Given the output of the cloth feature and human features, the AFE estimates the fused appearance flow. Subsequently, a residual connection is implemented to further enhance the quality of the generated appearance flow.

The structure of the FRW block is as follows: the image feature (whether cloth or human image) and the predicted appearance flow are processed through a spatial attention mechanism. This is followed by the fusion of the image feature maps and the appearance flow. A convolutional layer and a channel attention mechanism are subsequently introduced to rectify channel-related information. The ultimate step involves the application of a residual process to yield conclusive warping results.

\subsubsection{Cloth Generation Module}

Upon acquiring the flow and warping results, the cloth generation module is introduced to generate the virtual try-on images. This process entails several key steps. A shared encoder is deployed to map both the cloth image and the agnostic images into the latent space. Subsequently, the generated flow, warping outcomes, and one of the image features (either the cloth image or the agnostic image) are fused within the AFEW module. The aggregated features are then directed fed into a shallow decoder, ultimately generating the final virtual try-on result ($y_p$). Notably, the employed shallow encoder and decoder are each composed of two convolution layers without any down sampling.

\begin{figure*}[t!]
    \centering
    \includegraphics[width=.8\textwidth]{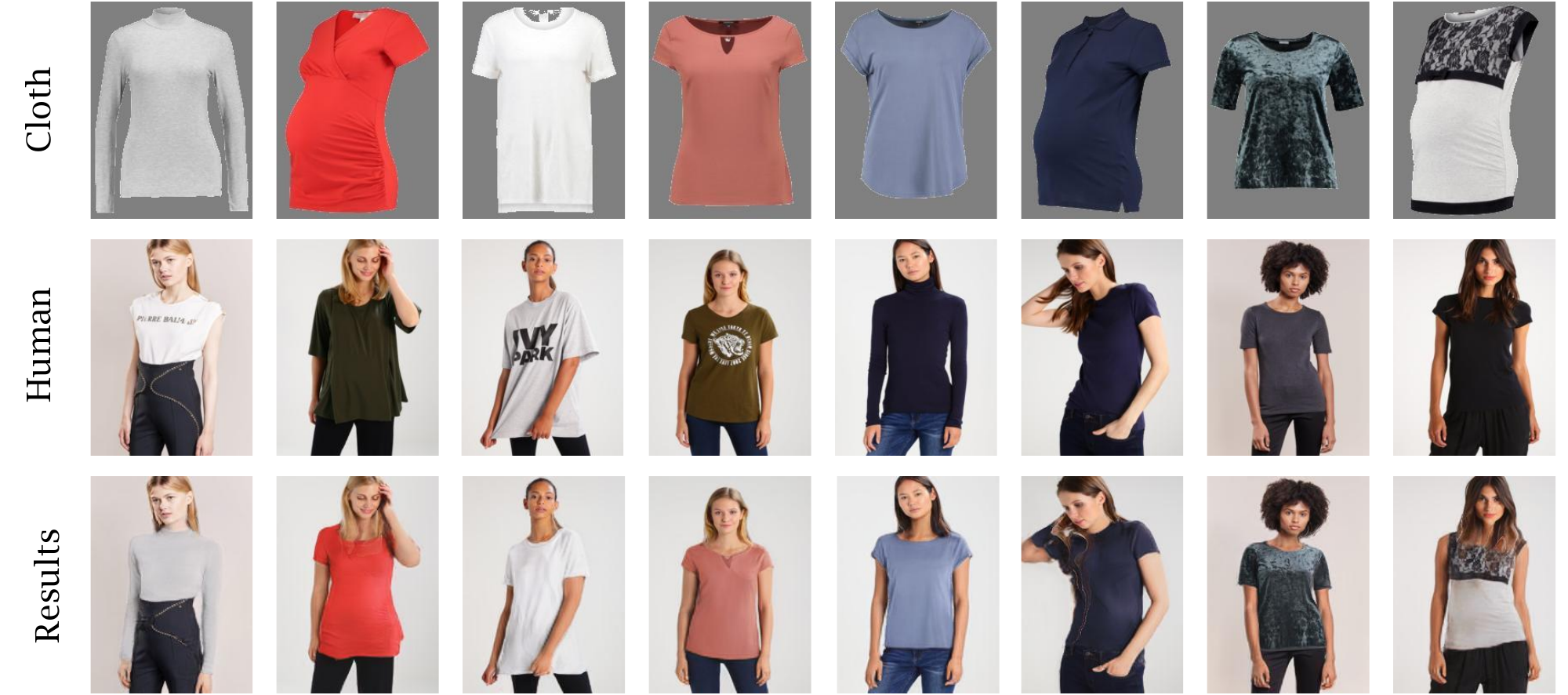}
    \caption{The visual results of the virtual try-on method on the VITON dataset.}
    \label{try_on}
\end{figure*}

\subsubsection{Training Losses}

The loss function has two primary components: the similarity loss function and the perceptual loss function~\cite{johnson2016perceptual}. The similarity loss is rooted in L1 distance and quantifies the dissimilarity between the predicted image and the corresponding ground truth image.

\begin{equation}
\mathcal{L}_{s}=|| y_p-y_g||_1
\label{eq:loss_test_all},
\end{equation}

Here, $||\cdot||_1$ denotes the L1 distance. The variables $y_p$ and $y_g$ respectively signify the predicted virtual try-on image and the ground truth image. The perceptual loss function is harnessed to ensure the consistency of feature maps, which are extracted through the VGG19 model~\cite{simonyan2014very}, between the predicted virtual try-on image and the ground truth image. This is mathematically formulated as follows:

\begin{equation}
\mathcal{L}{per}=\sum_{i=1}^5||\phi_i(\boldsymbol{y}_{p})-\phi_i(\boldsymbol{y}_{g})||_1
\label{eq:loss_test_all},
\end{equation}
where $\phi_i()$ signifies the features of the $i_{th}$ layer of the VGG-19 model. The comprehensive loss is presented as:

\begin{equation}
\mathcal{L}=\sum_{n=1}^N(n+1)*(\lambda_{s}\mathcal{L}_{s}^n+\lambda{per}\mathcal{L}_{per}^n)
\label{eq:loss_test_all},
\end{equation}
where, $\mathcal{L}^n$ symbolizes the loss at the $n_{th}$ scale.

\begin{table}[t!]
\renewcommand\arraystretch{1.5}
\caption{Quantitative comparisons with previous virtual try-on methods on the VITON dataset (paired setting).}
\centering
\setlength{\tabcolsep}{1.5mm}
\footnotesize
\begin{tabular}{ccccc}
\hline
Method  & FID ($\downarrow$) & SSIM  ($\uparrow$)& IS  ($\uparrow$)& PSNR ($\uparrow$)\\
\hline
CP-VTON &  30.50 & 0.784 & 2.757 & 21.01\\
ClothFlow &  23.68 & 0.843 & - & 23.60\\
ZFlow &  15.17 & 0.885 & - & 25.46\\
Ours &  11.32 & 0.887 & 2.846 & 26.19\\
\noalign{\smallskip}
\bottomrule
\end{tabular}
\vspace{-10pt}

\label{tab:comparative studies}
\end{table}

\section{Evaluation}
This section presents an evaluation of the MYCloth system's capabilities in two aspects: 1) virtual try-on performance, and 2) user experience. Firstly, we conduct a comparison experiment to validate the effectiveness of the virtual try-on model and an ablation study to validate the proposed modules. Then, we conducted a user study to determine if the MYCloth system can reduce discrepancies between the envisioned and final T-shirt. Finally, we conducted a user study to evaluate if the MYCloth system can improve the consumer's interactive experience when customizing clothes online.

\subsection{Evaluation of the virtual try-on}

\textbf{Implementation Details.} The cloth and human encoders within the warping module are both structured according to the Feature Pyramid Network (FPN)~\cite{lin2017feature}, comprising five downsampling stages. Notably, these encoders are independently parameterized. The appearance flow estimator, crucial for accurate warping, is constructed using four convolutional layers. Their hidden dimensions follow a sequence of 512, 256, 128, and 64, respectively.
In the cloth generation module, both the cloth and human encoders adopt an identical structure. This is reinforced by shared parameters. Each encoder and decoder is structured with three convolution layers, each with hidden dimensions of 32, 64, and 128. It's important to highlight that feature maps are not subsampled in this process.
For our experiments, we utilized the PyTorch framework on A6000 GPUs. Employing the Adam optimizer, we adopted a batch size of 16 for training. The training spanned 200 epochs, initialized with a learning rate of $5\times10^{-5}$. Subsequently, the learning rate underwent a reduction to 0.1 times the original every 50 epochs. The loss function weight factors, $\lambda_{s}$ and $\lambda_{per}$, were both set to 1, ensuring a balanced impact of the similarity and perceptual loss components.

\textbf{Datasets.} Our experimentation is grounded in the widely recognized VITON dataset~\cite{han2018viton}, a standard resource for virtual try-on research. This dataset serves as a pivotal benchmark in the field and comprises a training set encompassing 14,221 image pairs, alongside a distinct testing set containing 2,032 image pairs. Notably, each pair is composed of a front-view photograph and an in-shop clothing image, all calibrated to a resolution of 256 × 192 pixels. This dataset's established relevance and comprehensive coverage underscore its suitability for evaluating our proposed system.

\textbf{Comparison Experiments.} To evaluate the effectiveness of our method, we conducted a comparison experiment with previous virtual try-on methods, including Cloth-Flow~\cite{han2019clothflow}, CP-VTON~\cite{wang2018toward}, and ZFlow~\cite{chopra2021zflow}. In the comprehensive evaluation, we utilize several established metrics to assess the quality of our synthesized images compared to ground truth images. These metrics include the Structure Similarity Index Measure (SSIM)~\cite{seshadrinathan2008unifying}, the Peak Signal-to-Noise Ratio (PSNR)~\cite{hore2010image}, the Fréchet Inception Distance (FID)~\cite{heusel2017gans}, and the Inception Score (IS)~\cite{salimans2016improved}. These metrics collectively measure structural similarity, image fidelity, distributional similarity, and the realism of the generated images. Ground truth images of the same individuals wearing identical clothing are employed as references for these evaluations, ensuring a robust and quantitative assessment of our system's performance.

In our comprehensive quantitative evaluation of virtual try-on methods using the VITON dataset, as shown in Tab.~\ref{tab:comparative studies}, our proposed method demonstrates exceptional performance across multiple key metrics. With the lowest FID of 11.32, the SSIM of 0.887, a competitive IS of 2.846, and the highest PSNR of 26.19, our method consistently outperforms other previous approaches, such as CP-VTON, ClothFlow, and ZFlow. These results underscore the effectiveness of our method in producing virtual try-on images that are not only highly realistic and structurally similar to real images but also diverse in content, making it a compelling choice for enhancing the online customized shopping experience.

To evaluate the efficacy of our algorithm, we conducted an empirical study employing the VITON dataset. The visual results of this study are elucidated in Fig.~\ref{try_on}. In this figure, we observe three columns of images. The first column showcases the reference cloth images, while the second column exhibits the human images. The third column presents the outcomes achieved through the application of our method. Notably, our algorithm demonstrates remarkable proficiency in capturing intricate textures and faithfully reproducing poses. 

\textbf{Ablation Experiments.} 
In Tab.~\ref{tab:ablation study}, we systematically analyzed the impact of different modules on our framework's performance, using abbreviated notations for clarity. We started with the baseline configuration and progressively introduced components. The Attention-based Flow Estimation and Warping Module (AFEW) significantly improved both SSIM and PSNR, highlighting its value in enhancing image quality. The incorporation of the Flaw Rectification and Warping Block (FRW) within the Warping module and later within the cloth generation module further elevated these metrics, indicating their crucial roles in refining the warping and generation processes. Ultimately, the combined presence of all modules led to the highest SSIM (0.887) and PSNR (26.19) values, showcasing the effectiveness of these components and underscoring their collective contribution to superior performance.

\begin{table*}[t!]
\renewcommand\arraystretch{1.5}
\caption{Quantitative comparisons with previous virtual try-on methods on the VITON dataset (paired setting).}
\centering
\setlength{\tabcolsep}{1.5mm}
\footnotesize
\begin{tabular}{cccccc}
\hline
Baseline  & AFEW & FRW within the Warping module & FRW within the cloth generation module & SSIM  ($\uparrow$)& PSNR ($\uparrow$)\\
\hline
\checkmark &  & & & 0.815 & 20.26\\
\checkmark & \checkmark &  & & 0.862 & 24.73\\
\checkmark &  & \checkmark & & 0.828 & 21.95\\
\checkmark & \checkmark & \checkmark & & 0.871 & 25.95\\
\checkmark & \checkmark & \checkmark & \checkmark & 0.887 & 26.19\\
\noalign{\smallskip}\hline
\end{tabular}

\label{tab:ablation study}
\end{table*}

\subsection{Evaluation of the User Experience}
To assess the practicality of our system, we conducted comprehensive evaluations involving 15 students and 3 T-shirt customization sellers. Data collection was accomplished through a combination of quantitative analysis, involving subjective ratings, and qualitative research via semi-structured interviews, with the overarching goal of thoroughly assessing the user experience.

\begin{figure}[t!]
    \centering
    \includegraphics[width=.5\textwidth]{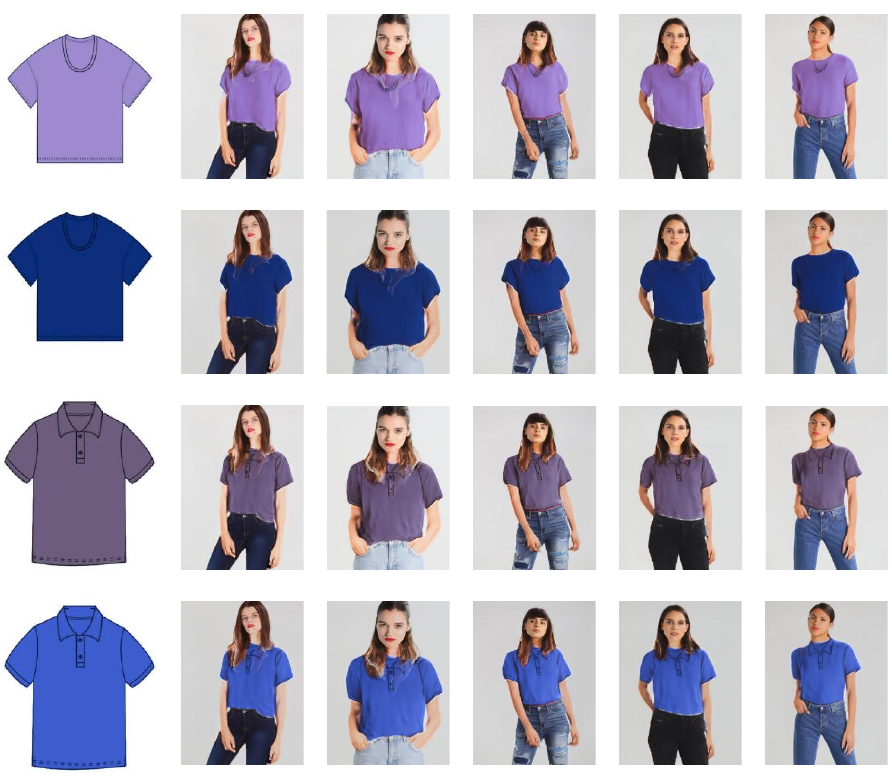}
    \caption{Example of the user uses our system to change the color of a cloth in a chosen pattern and uses the virtual try-on to see how it looks on the body..}
    \label{responses_1}
\end{figure}

\begin{figure}[t!]
    \centering
    \includegraphics[width=.5\textwidth]{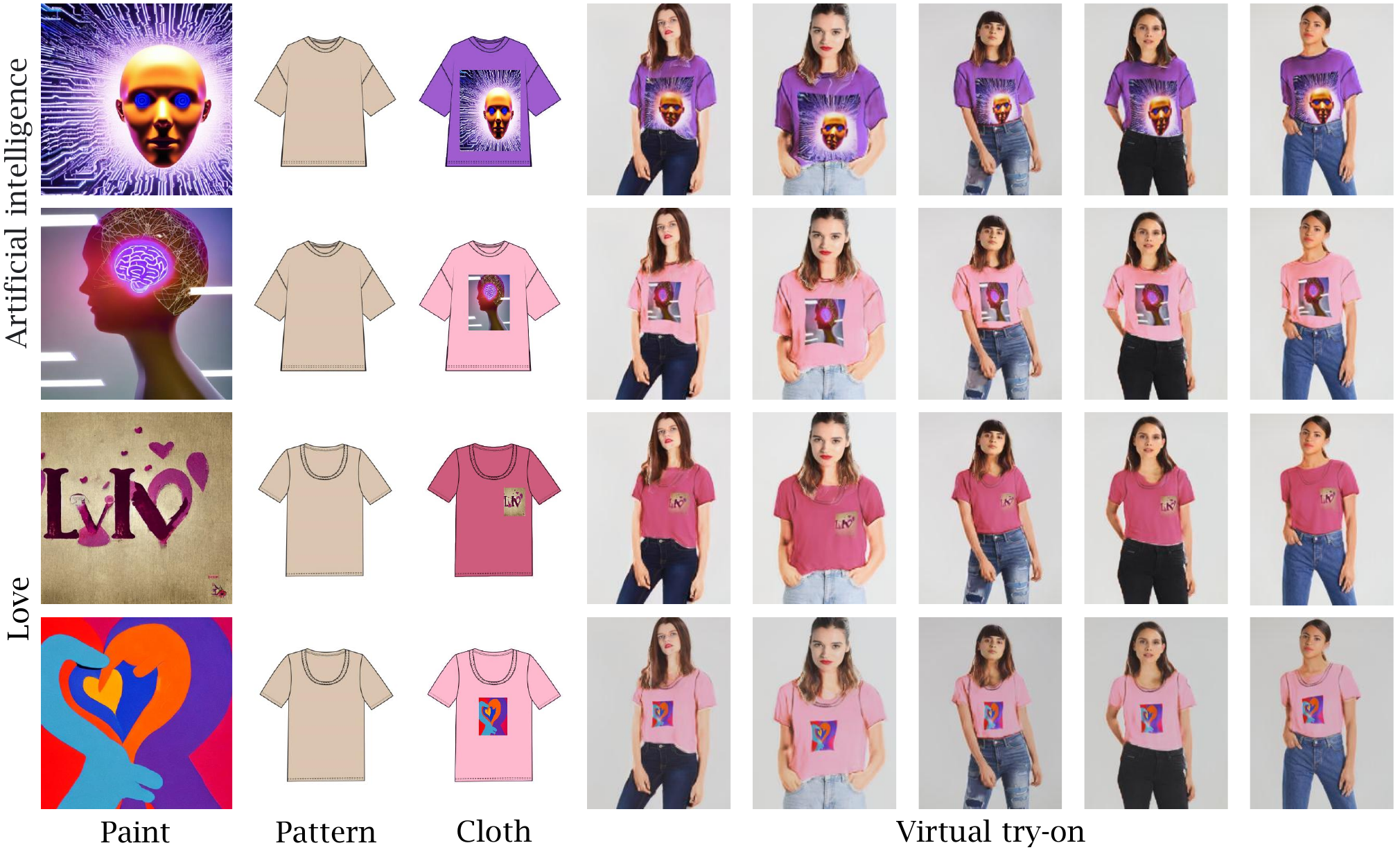}
    \caption{Users use our system to generate prints based on a given theme, change the color of the T-shirt, change the size and position of the prints, and use the virtual try-on to see an example of how it looks on the body.}
    \label{responses_2}
\end{figure}

\subsubsection{Study Design}

We structured our evaluation approach with tailored surveys for consumers and T-shirt customization sellers. Each participant was invited to use our system to create a customized T-shirt item based on a specific theme, followed by individual interviews to gather insights.

\subsubsection{Participants}
Among the participants, 8 participants were female and 7 participants were male, they were between 18 and 27 years old. They represented diverse academic backgrounds, including public policy (4 participants), computer artificial intelligence (3 participants), mechanical engineering (2 participants), applied mathematics (2 participants), bioengineering (2 participants), and finance (2 participants). They all had experience wearing group-customized T-shirts. For the T-shirt customization sellers, we engaged three online T-shirt customization sellers based on their established sales.

\subsubsection{Task and procedures}
Each participant was tasked with completing four specific assignments:

\begin{itemize}
  \item \textbf{Style and Color Selection.} Participants are required to select a T-shirt style as the starting point for their personalized design. The objective is to establish the foundational elements of the customized T-shirt design.
  \item \textbf{Paint Generation.} Participants input a theme or text that will guide their T-shirt design. They can enter slogans, quotes, or any text they wish to feature on the T-shirt. This text provides a personalized touch to the design, aligning it with their creative vision.
  \item \textbf{Cloth Editing.} This task allows participants to edit and fine-tune the pattern elements of their T-shirt design. They can change the T-shirt's color, adjust print sizes and positions, and even add additional pattern elements like icons or shapes. 
  \item \textbf{Virtual Try-On.} Participants virtually try on their personalized T-shirt designs. This step provides insights into the final look and fit of the customized T-shirt.
  
\end{itemize}
\subsubsection{Measurement}
After participants completed their tasks, they were asked to provide assessments on various aspects of the system using a 7-point Likert scale. 

\subsubsection{Results: overall assessment}
\begin{figure}[t!]
    \centering
    \includegraphics[width=.5\textwidth]{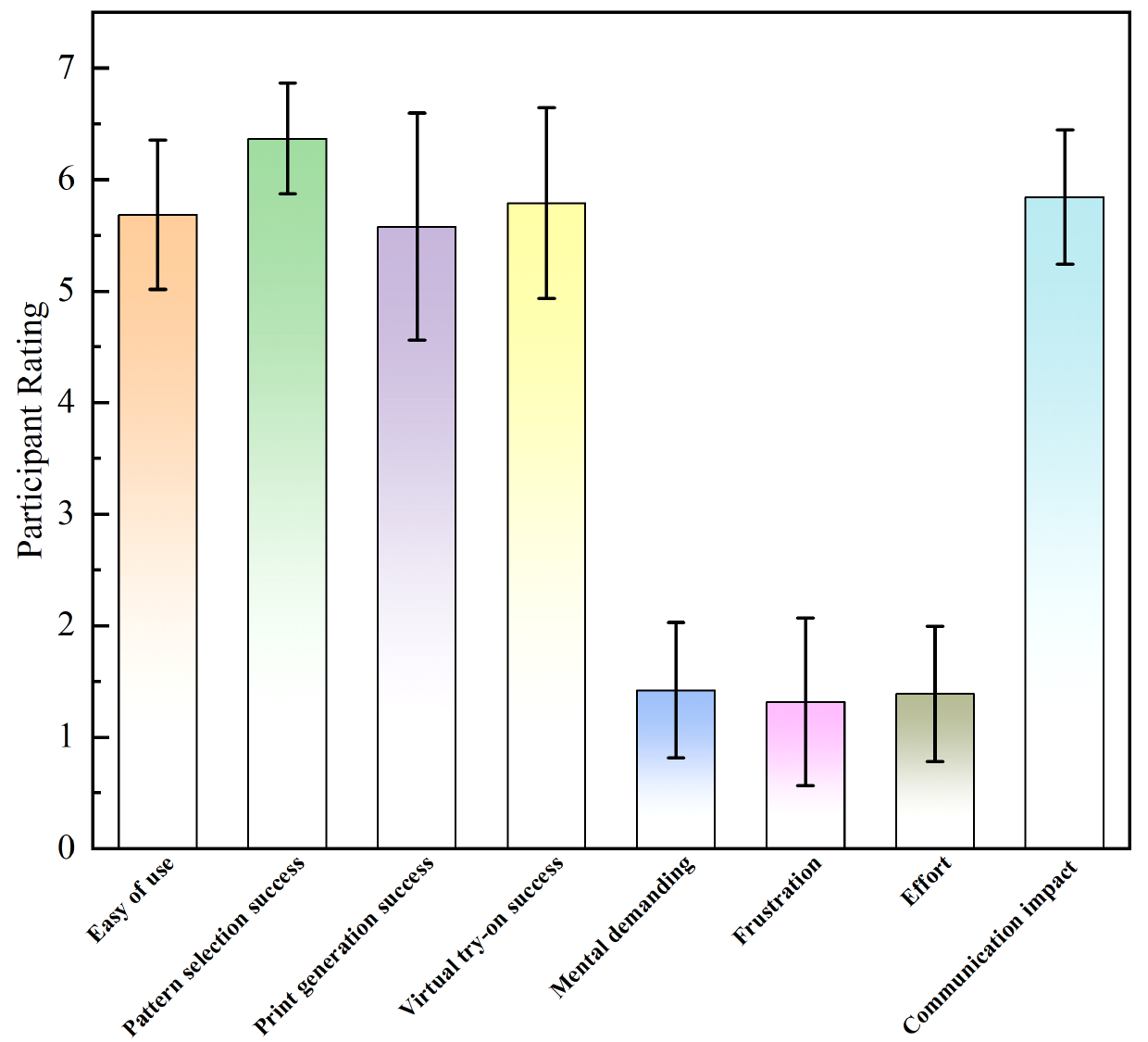}
    \caption{The responses from participants were evaluated using a seven-point Likert scale for all questions.}
    \label{responses}
\end{figure}

\begin{figure}[t!]
    \centering
    \includegraphics[width=.5\textwidth]{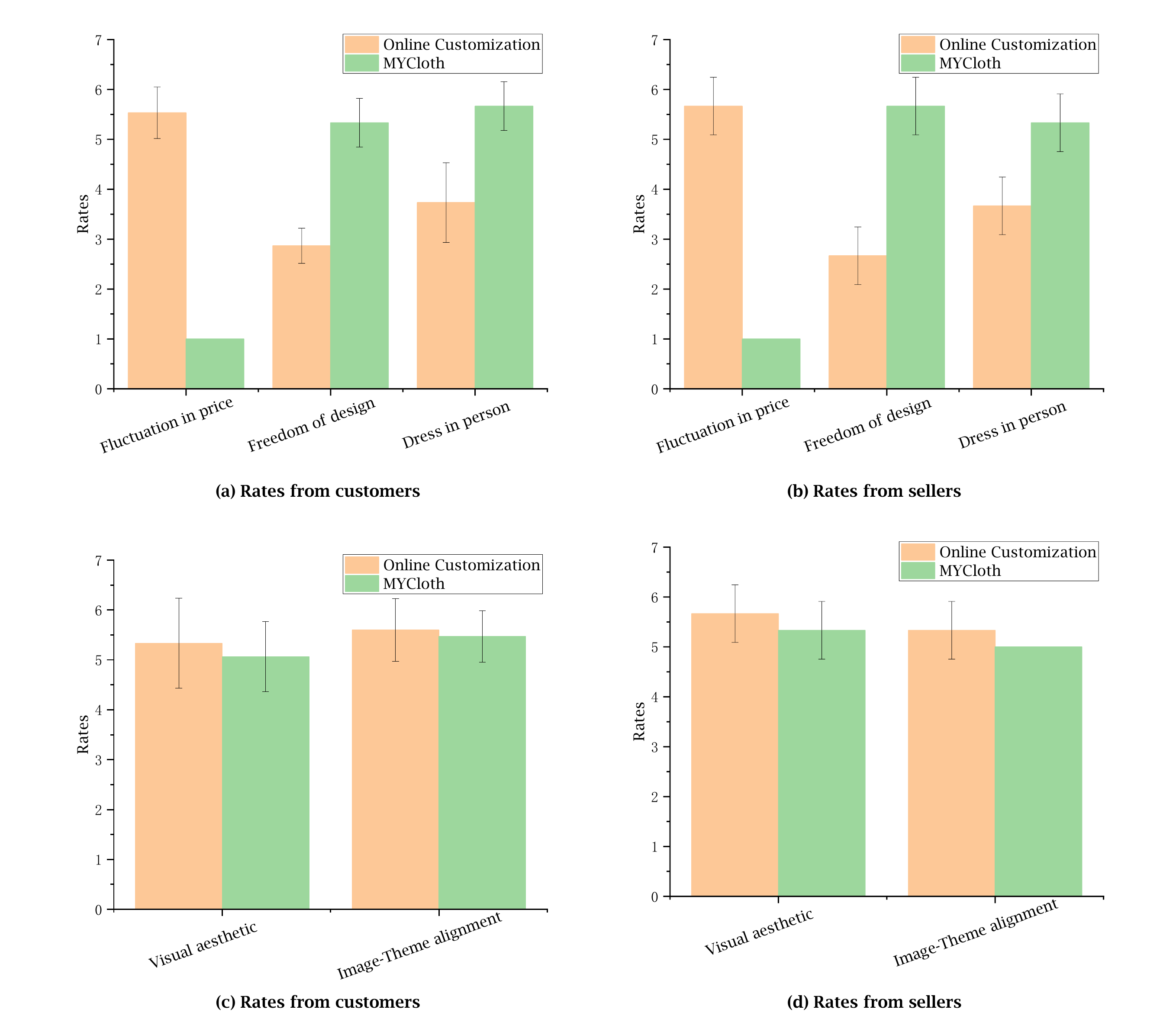}
    \caption{The responses from customers and sellers were evaluated using a seven-point Likert scale.}
    \label{Rates}
\end{figure}

In this section, we provide an overview of the questionnaire responses collected from users who engaged with the MYCloth system, as depicted in Fig.~\ref{responses} and Fig.~\ref{Rates}. Participants are identified by labels ranging from P1 to P18, with P1-15 representing consumers and P16-18 representing T-shirt customization sellers. The ratings encompass various aspects, including system usability, pattern selection satisfaction, print generation success, virtual try-on effectiveness, mental demand, frustration, effort, and communication impact. Examples obtained by some users using our system are shown in Fig.~\ref{responses_1} and Fig.~\ref{responses_2}.

\section{Conclusion}
In this paper, we introduced MYCloth, an innovative online T-shirt customization system designed to enhance personalized fashion experience, offering users the opportunity to become designers themselves and create a T-shirt that aligns with their unique tastes. With an intuitive interface, advanced print generation based on a large language model and stable diffusion model, and a robust virtual try-on model, MYCloth redefines the customization process. Through extensive user surveys, interviews, and usability assessments involving both consumers and T-shirt customization sellers, this research has underscored the system's efficacy. Users consistently expressed high satisfaction with MYCloth's user-friendliness, the realism of virtual try-ons, and the quality of generated prints. These results emphasize how MYCloth enhances T-shirt customization, offering an enjoyable, efficient, and unique way to personalize fashion choices.



{
    \small
    \bibliographystyle{ieeetr}
    \bibliography{ref}
}

\end{document}